\documentclass[conference]{IEEEtran}

\ifCLASSINFOpdf
\else
\fi
\usepackage[utf8]{inputenc}
\usepackage{amsmath,amssymb,amsfonts}
\usepackage{cite}

\usepackage{algorithmic}

\usepackage{graphicx}
\usepackage{textcomp}
\usepackage{xcolor}
\usepackage{pifont}
\usepackage{amsthm}
\usepackage{mathrsfs}
\usepackage{cases}
\usepackage{comment}
\usepackage{empheq} 
\usepackage{caption}
\usepackage{subcaption}

\allowdisplaybreaks

\usepackage{url}  
\usepackage{graphicx}  
\usepackage[ruled,vlined,linesnumbered]{algorithm2e}
\usepackage{setspace}
\usepackage{floatpag} 
\usepackage{float}
\floatpagestyle{empty}

\usepackage[top=0.5in, bottom=0.5in, left=0.5in, right=0.5in]{geometry}

\usepackage{listings}

\usepackage{fancyhdr}

\begin{document}
\title{IRS-Assisted Millimeter Wave Communications: Joint Power Allocation and Beamforming Design}
%
%
%

\author{\IEEEauthorblockN{Yue Xiu\textsuperscript{1,3},~Yang Zhao\textsuperscript{1},~\textup{Yang Liu}$^{1}$,~Jun Zhao\textsuperscript{1},~Osman Yagan\textsuperscript{2},~Ning Wei\textsuperscript{3}}
\IEEEauthorblockA{\textsuperscript{1}School of Computer Science and Engineering,
Nanyang Technological University, Singapore 639798\\
\textsuperscript{2}Carnegie Mellon University, Pittsburgh, PA 15213, United States\\
\textsuperscript{3}University of Electronic Science and Technology of China, Chengdu, China, 611731 \\
xiuyue@std.uestc.edu.cn, \{yang.zhao, yang-liu, junzhao\}@ntu.edu.sg, oyagan@ece.cmu.edu, wn@uestc.edu.cn\\
}
}

\renewcommand{\headrulewidth}{0pt}

\maketitle

\begin{abstract}
Intelligent reflecting surface (IRS) technology offers more feasible propagation paths for millimeter-wave (mmWave) communication systems to overcome blockage than existing technologies. In this paper, we consider a downlink wireless system with the IRS and formulate a joint power allocation and beamforming design problem to maximize the weighted sum-rate, which is a multi-variable optimization problem.  To solve the problem, we propose a novel alternating manifold optimization based beamforming algorithm. Simulation results show that our proposed optimization algorithm outperforms existing algorithms significantly in improving the weighted sum-rate of the wireless communication system.
\end{abstract}


%
\IEEEpeerreviewmaketitle

\section{Introduction}
\IEEEPARstart{M}{illimeter}-wave (mmWave) communication is an essential technology for 5G, which addresses the bandwidth shortage problem in current mobile systems~\cite{alkhateeb2015limited}. However, communications using the mmWave bands suffer from a higher path loss than communications with the low-frequency bands~\cite{wang2015multi}. Recently, the intelligent reflecting surface (IRS) has been  proposed as a promising technology to solve the above challenges~\cite{wu2019beamforming, huang2019reconfigurable}.  A key problem for IRS-assisted systems is to optimize the transmit beamforming which is called the active beamforming at the BS and reflect beamforming which is also named as the passive beamforming at the IRS to maximize the weighted sum-rate. Thus, we propose an alternative optimization algorithm to maximize the weighted sum-rate.

%
%
%
%


Most existing studies on beamforming design of the IRS overcome the difficulty of multi-variable optimization problem by decoupling the original problem into active and passive beamforming sub-problems and then focus on the constraints in solving the sub-problems~\cite{xie2019max,xia2019intelligent,
zheng2019intelligent,
chen2019intelligent,
guo2019weighted,
yan2019passive, wu2019beamforming, wang2019channel
}. Under the popular model~\cite{wu2019intelligent} that the passive beamforming at the IRS includes only phase shifts, the IRS passive beamforming is difficult to optimize  due to the \mbox{non-convex} unimodular constraints of its elements. One effective and widely used approach is to consider the beamforming optimization on the manifold~\cite{lu2017linear,yu2016alternating,chen2017low,lin2019hybrid, yu2019miso}. Therefore, we consider the manifold optimization to resolve the optimization problem with unimodular constraints. In addition, by recalling the work in mmWave systems~\cite{zhu2019millimeter, zhu2019millimeterwave}, the joint power allocation and beamforming design is another important problem. One direct motivation to consider the problem is that a practical system is constrained to the transmitted power~\cite{zhu2019millimeterwave}.  Therefore,  we maximize the weighted sum-rate by optimizing the power allocation matrix and active/passive matrix for the IRS-assisted system.

The contributions of this paper can be summarized as follows: we formulate the power allocation problem and solve the optimization problem by using geometric programming (GP)~\cite{chiang2007power}.  In the beamforming design stage, we consider the unimodular constraints of the phase shifts and unit-vector constraints of the normalized active beamforming as Riemannian manifold~\cite{absil2009optimization} and Oblique manifold~\cite{absil2009optimization}, respectively. The approaches for optimization on two manifolds are relatively limited. By using the alternating optimization method,  we derive the gradient of the objective function on two manifolds. Then, we propose to use the conjugate gradient method to search the optimal active/passive beamforming. Moreover, the convergence analysis for the proposed approach is provided. Furthermore, we apply the random beamforming algorithm in~\cite{lee2015randomly} to the IRS-assisted system and set its performance as the benchmark.

The rest of the paper is organized as follows. Section~\ref{sec-related} surveys the related literature on the manifold and IRS, while in Section~\ref{III}, we introduce the system model. We present the problem formulation in Section~\ref{IV}. In Section~\ref{V}, we offer a joint power allocation and beamforming design algorithm and analyze the convergence and computational complexity of the proposed algorithm. We demonstrate various simulation results in Section~\ref{VI}.  The conclusion is drawn in Section~\ref{VII}.
\section{Related Work}~\label{sec-related}
Multi-variable optimizations have been widely used in IRS-assisted beamforming designs, and some solutions have been proposed to solve the problem where one effective and popular approach is to consider the beamforming design on the manifold~\cite{guo2019weighted,xu2019resource,yu2019miso}. For example, Guo~\emph{et~al.}~\cite{guo2019weighted} investigated the IRS-aided multiuser downlink multi-input single-output (MISO) system and proposed an algorithm on the manifold to solve the IRS phase optimization problem for a joint transmit beamforming design and the IRS phase optimization problem to maximize the weighted sum-rate under the AP transmit power constraint; Yu~\emph{et~al.}~\cite{yu2019miso} investigated the joint design of the beamformer for the IRS-assisted system and proposed a manifold optimization based algorithm; Cao~\emph{et~al.}~\cite{cao2020intelligent} proposed a novel manifold alternative optimization algorithm to minimize the uplink transmit power for all users; Xu~\emph{et~al.}~\cite{xu2019resource} proposed a manifold optimization based low-complexity beamforming algorithm for the IRS-assisted security communication system.  Although the above studies design the passive and active beamforming with the objective of maximizing the sum-rate, they only optimize the passive beamforming on one manifold. \cite{pan2020multicell,
li2020irs,
mu2020exploiting,
zhou2020intelligent,
pan2020intelligent} considered the weighted sum-rate maximization problem for various IRS-aided systems. The authors in \cite{pan2020multicell,pan2020intelligent} studied the weighted sum-rate maximization problem for IRS-assisted simultaneous wireless information and
power transfer (SWIPT) system and multicell system. In \cite{li2020irs,
mu2020exploiting}, the authors studied the IRS-assisted the weighted sum-rate maximization problem for orthogonal frequency division multiplexing (OFDM) system and non-orthogonal multiple access (NOMA) system. In \cite{zhou2020intelligent}, the authors studied the weighted sum-rate maximization problem for IRS-assisted Multigroup
Multicast MISO communications. However, the above papers did not consider the weighted sum-rate maximization problem for mmWave systems. Although \cite{cao2020intelligent} considered the weighted sum-rate maximization for mmWave system, the authors did not consider the mmWave channel without NLOS. Therefore, in this paper, we consider the joint power allocation and beamforming design problem of the IRS-assisted mmWave system without NLOS.  In this problem, except for the difficulty of the joint optimization over the power variable and two beamforming variables (power allocation matrix and active/passive beamformers), the unimodular constraints of the passive beamformers due to the controller make the problem highly non-convex and difficult to solve~\cite{guo2018broad}. Hence, we investigate the joint power allocation and passive/active beamforming optimization for IRS-assisted multi-user mmWave systems, aiming at maximizing the weighted sum rate. In addition, using manifold optimization technology is able to resolve problems in the optimization of the two beamforming variables with the unit modulus and the unit sum constraint. Hence, we propose the alternating optimization algorithms based on GP~\cite{chiang2007power} and manifold optimization~\cite{absil2009optimization} for IRS-assisted multi-user mmWave systems to maximize the weighted sum-rate.

In IRS-assisted systems with mmWave, in addition to sum-rate optimization of multiple users addressed in our paper and~\cite{guo2019weighted,cao2020intelligent, di2020hybrid}, many other problems have also been investigated in the literature, including information rate maximization of one user in~\cite{perovic2019channel,wang2019intelligent,wang2019joint}, channel estimation in~\cite{wang2019compressed,taha2019deep,taha2019enabling, zheng2020fast, you2020channel}, positioning in~\cite{he2019large,he2019adaptive,zhang2020towards}, optimization of the number of phase shifts~\cite{zhang2020reconfigurable}, indoor localization~\cite{zhang2020metalocalization}, and maximization of the secured transmission~\cite{li2020reconfigurable}. \vspace{-10pt}

\section{System Model}\label{III}
\begin{figure}[!t]
\centering
\includegraphics[height=1.5in]{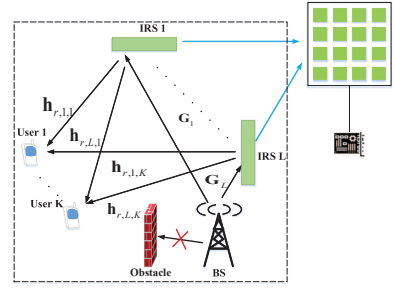}
\caption{IRS-assisted mmWave downlink multi-user communication system.\vspace{-20pt}}
\label{fig-system}
\end{figure}

In this paper, we consider an IRS-assisted mmWave system shown in Fig.~\ref{fig-system}. The base station (BS) is equipped with an array of $N$ antennas and serves $K$ single-antenna users. $L$ IRSs are deployed to assist the data transmission from the BS to users, where each IRS is assumed to include $M_{x}$ antennas horizontally and $M_{y}$ antennas vertically. Thus, the total number of antennas at each IRS is $M=M_{x}\times M_{y}$. Since mmWave links are highly susceptible to blockages, we neglect the direct link between the BS and each user.

Let $\mathbf{G}_{l}\in\mathbb{C}^{M\times N}$ denotes the mmWave channel matrix for the channel between BS and the $l$-th IRS; $\mathbf{h}_{r,l,k}\in\mathbb{C}^{M\times 1}$ denotes the channel between the $l$-th IRS and the $k$-th user. The phase shift matrix  $\boldsymbol{\Theta}_{l}$ of the $l$-th IRS is defined as
\begin{eqnarray}
\boldsymbol{\Theta}_{l}&=&\mathrm{diag}(\boldsymbol{\theta}_{l}),
\end{eqnarray}
\text{where}~$\boldsymbol{\theta}_{l}=
[e^{j\theta_{l,1}},\cdots,e^{j\theta_{l,m}},\cdots,e^{j\theta_{l,M}}]^{\mathrm{T}}$.

The signal received at the $k$-th user can be written as
\begin{eqnarray}
y_{k}&=&\underbrace{\sum_{l=1}^{L}\mathbf{h}_{r,l,k}^{H}\boldsymbol{\Theta}_{l}\mathbf{G}_{l}\mathbf{w}_{k}p_{k}s_{k}}_{\textrm{desired information signal}}\nonumber\\
&\quad&\quad\quad\quad+\underbrace{\sum_{i\neq k}^{K}\sum_{l=1}^{L}\mathbf{h}_{r,l,k}^{\mathrm{H}}\boldsymbol{\Theta}_{l}\mathbf{G}_{l}\mathbf{w}_{i}p_{i}s_{i}}_{\textrm{interference signal}}+n_{k},\label{2}
\end{eqnarray}
where $\mathbf{W}=[\mathbf{w}_{1},\cdots,\mathbf{w}_{K}]^{H}\in\mathbb{C}^{K\times N}$ is the active beamforming matrix, and $\mathbf{w}_{k}\in\mathbb{C}^{N\times 1}$ is used by the BS to transmit the symbol $s_{k}$. $\mathbf{P}\in\mathbb{C}^{K\times K}$ is the power allocation matrix and $\mathbf{P}=\mathrm{diag}(\mathbf{p})$, where $\mathbf{p}=[p_{1},\cdots,p_{K}]\in\mathbb{C}^{K\times 1}$. $n_{k}\sim\mathcal{CN}(0,\sigma^{2})$ represents the circularly symmetric complex Gaussian (CSGS) noise with mean $0$ and variance $\sigma^{2}$. $\mathbf{s}=[s_{1},\cdots,s_{K}]^{T}\in\mathbb{C}^{K\times 1}$ is the transmit signal, where $s_{k}$ is the signal from the BS to the $k$-th user with mean $0$ and normalized power of $\mathbb{E}[|s_{k}|^{2}]=1$.

Since we separate the transmission power from beamforming matrix $\mathbf{W}$, it is without loss of generality to assume that each row of $\mathbf{W}$ has a unit norm, i.e.,
\begin{equation}
\begin{split}
\|\mathbf{w}_{k}\|^{2}=1,~\forall 1\leq k \leq K.
\end{split}
\end{equation}

We adopt the widely used rank-one structure channel model~\cite{wang2019compressed, li2019joint} and use the directional mmWave channel with a uniform linear array (ULA) with $d=\frac{\lambda}{2}$ antenna spacing, where   $\lambda$ is the wavelength. The channel matrix is denoted as
\begin{equation}
\begin{split}
\mathbf{G}_{l}=\gamma_{l}\mathbf{a}_{r}(\phi_{r}^{l},\theta_{r}^{l})\mathbf{a}_{t}^{H}(\phi_{t}^{l}),~\forall l,
\end{split}
\end{equation}
where $\gamma_{l}$ represents the complex channel gain of the $l$-th IRS to BS. $\phi_{r}^{l}$ and $\theta_{r}^{l}$ denote the elevation and azimuth angles of IRS, respectively. The array steering vector $\mathbf{a}_{r}(\phi_{r}^{l},\theta_{r}^{l})$  is defined as
\begin{equation}
\begin{split}
\mathbf{a}_{r}(\phi_{r}^{l},\theta_{r}^{l})=\mathbf{a}_{r}^{az}(\theta_{r}^{l})\otimes\mathbf{a}_{r}^{el}(\phi_{r}^{l}),
\end{split}
\end{equation}
where $\mathbf{a}_{r}^{az}(\theta_{r}^{l})$ and $\mathbf{a}_{r}^{el}(\phi_{r}^{l})$ are the horizontal steering vector and the vertical steering vector of the IRS.

Similarly, the channel between the IRS and each user is also line-of-sight (LOS) dominated and has a rank-one structure, and the mmWave channel $\mathbf{h}_{r,l,k}$ can be denoted as
\begin{equation}
\begin{split}
\mathbf{h}_{r,l,k}=\rho_{l,k}\mathbf{a}_{t}(\phi_{t}^{l,k}),
\end{split}
\end{equation}
where $\rho_{l,k}$ is the channel gain.

\section{Problem Formulation}~\label{IV}
According to (\ref{2}), the achievable rate of the $k$-th user can be formulated as
\begin{equation}
\begin{split}
R_{k}&=\log_{2}\left(1+\frac{|\sum_{l=1}^{L}\mathbf{h}_{r,l,k}^{\mathrm{H}}\boldsymbol{\Theta}_{l}\mathbf{G}_{l}\mathbf{w}_{k}|^{2}p_{k}}{\sum_{i\neq k}^{K}|\sum_{l=1}^{L}\mathbf{h}_{r,l,k}^{\mathrm{H}}\boldsymbol{\Theta}_{l}\mathbf{G}_{l}\mathbf{w}_{i}|^{2}p_{i}+\sigma^{2}}\right).
\end{split}
\end{equation}
The optimization problem to maximize the weighted sum rate is formulated as follows, where $\omega_{k}|_{k\in\{1,\ldots,K\}}$ denote the weights: 
\begin{subequations}
\begin{align}
\max_{\boldsymbol{\theta},\{\mathbf{w}_{k}\}_{k=1}^{K},\mathbf{p}}\quad & \sum_{k=1}^{K}\omega_{k}R_{k}\\
\mbox{s.t.}\quad
&C_{1}:\sum_{k=1}^{K}p_{k}\leq P,~~~~~p_{k}> 0,~ \forall k,&\\
&C_{2}:|\theta_{i}|=1,~\forall i, &\\
&C_{3}:\|\mathbf{w}_{k}\|^{2}=1,~\forall k, &
\end{align}
\label{8} 
\end{subequations}
where constraint $C_{1}$ imposes the maximum total power budget $P$. $C_{2}$ is the unit modulus constraint for the passive beamforming. $C_{3}$ denotes the  amplitude constraint of each row in the active beamforming matrix.

\section{Joint Power Allocation and Beamforming Design}~\label{V}
As we can see from~\eqref{8}, the main difficulty of this problem is that the optimized variables are entangled with each other. Instead of solving the three variables optimization problem, we propose a two-stage approach in this section.  The two stages consist of power allocation and beamforming design.  For the first stage, the power allocation problem with the fixed beamforming matrices is expressed as
\begin{subequations}
\begin{alignat}{2}
\max_{\mathbf{p}}\quad & f(\mathbf{p})~\label{9a}\\
\mbox{s.t.}\quad
&C_{1}:\sum_{k=1}^{K}p_{k}\leq P.&
\end{alignat}\label{9}
\end{subequations} 
We can solve the problem~(\ref{9}) using the GP algorithm~\cite{chiang2007power}. The simple GP algorithm's details are omitted here, whereas we mainly focus on the beamforming design algorithm. When the power allocation matrix $\mathbf{P}$ is fixed, the optimization problem~(\ref{8}) can be simplified as
\begin{subequations}
\begin{alignat}{2}
\max_{\boldsymbol{\theta},\mathbf{w}_{1},\cdots,\mathbf{w}_{K}}\quad & f(\boldsymbol{\theta},\mathbf{w}_{1},\cdots,\mathbf{w}_{K})\\
\mbox{s.t.}\quad
&C_{1}:|\theta_{i}|=1,\forall i, &\\
&C_{2}:\|\mathbf{w}_{k}\|^{2}=1,\forall k, &\label{10c}
\end{alignat}~\label{10}
\end{subequations}
where $f(\boldsymbol{\theta},\mathbf{w}_{1},\cdots,\mathbf{w}_{K})=\sum_{k=1}^{K}w_{k}R_{k}$.
To solve the problem~(\ref{10}), we propose a manifold optimization based beamforming algorithm.

\subsection{Alternating Manifold Optimization Algorithm for Beamforming}
It is not difficult to find the fact that constraints of the problem~(\ref{10}) can be viewed as Riemannian manifold~\cite{absil2009optimization} and Oblique manifold~\cite{absil2009optimization}. Therefore, we consider~(\ref{10}) as an optimization problem over the manifold space. Although the optimiation problems on a single manifold have been investigated in~\cite{lu2017linear,yu2016alternating}, the optimization problems on two manifolds are  rarelyconsidered. In this subsection, we propose one alternative optimization algorithm over two manifold spaces. 

Specifically, the active beamforming matrix constraints are $\|\mathbf{w}_{k}\|^{2}= 1,\forall k$. Since $\|\mathbf{w}_{k}\|^{2}= 1,\forall k$ is not a standard Oblique manifold~\cite{absil2009optimization}, we need to transform the optimization constraint (\ref{10c}) to the standard Oblique manifold. Hence, instead of optimizing the $\{\mathbf{w}_{k}\}_{k=1}^{K}$, we optimize   $\mathbf{W}$, and the constraint condition is transformed as
\begin{equation}
\begin{split}
\|\mathbf{w}_{k}\|^{2}= 1\Longleftrightarrow\mathbf{I}_{K\times K}\circ(\mathbf{W}\mathbf{W}^{H})=\mathbf{I}_{K\times K},
\end{split}
\end{equation}
where $\mathbf{I}_{K\times K}$   is an $K\times K$ identity matrix. Problem (\ref{10}) is rewritten as
\begin{subequations}
\begin{alignat}{2}
\max_{\boldsymbol{\theta},\mathbf{W}}\quad & f(\boldsymbol{\theta},\mathbf{W})\\
\mbox{s.t.}\quad
&C_{1}:|\theta_{i}|=1, \forall i,&\label{12b}\\
&C_{2}:\mathbf{I}_{K\times K}\circ(\mathbf{W}\mathbf{W}^{H})=\mathbf{I}_{K\times K}\label{12c}.
\end{alignat}\label{12}
\end{subequations}

To optimize the passive beamforming $\boldsymbol{\theta}$ with fixed active beamforming $\mathbf{W}$, problem~(\ref{12}) is written as
\begin{subequations}
\begin{alignat}{2}
\max_{\boldsymbol{\theta}}\quad & f_{1}(\boldsymbol{\theta})\\
\mbox{s.t.}\quad
&C_{1}:|\theta_{i}|=1, \forall i,&
\end{alignat}
\end{subequations}
where $f_{1}(\boldsymbol{\theta})=\sum_{k=1}^{K}w_{k}R_{k}$, and $\{\mathbf{w}_{k}\}_{k=1}^{K}$ are fixed values. Similarly, when designing $\mathbf{W}$, (\ref{12}) can be expressed as
\begin{subequations}
\begin{alignat}{2}
\max_{\mathbf{W}}\quad & f_{2}(\mathbf{W})\\
\mbox{s.t.}\quad
&C_{1}:\mathbf{I}_{K\times K}\circ(\mathbf{W}\mathbf{W}^{H})=\mathbf{I}_{K\times K},
\end{alignat}
\end{subequations}
where $f_{2}(\mathbf{W})=\sum_{k=1}^{K}w_{k}R_{k}$, and $\boldsymbol{\theta}$ is a fixed value. Next, we calculate the gradients and the projections on the two manifolds.

Let the Euclidean gradients of the weighted sum rate $f_{1}(\boldsymbol{\theta})$ over $\boldsymbol{\theta}$ and $f_{2}(\mathbf{W})$ over $\mathbf{W}$ be respectively defined as
\begin{align}
\begin{split}
\nabla f_{1}(\boldsymbol{\theta})&=\frac{\partial f_{1}(\boldsymbol{\theta})}{\partial
\boldsymbol{\theta}^{*}},\quad\nabla f_{2}(\mathbf{W})=\frac{\partial f_{2}(\mathbf{W})}{\partial \mathbf{W}^{*}}.
\end{split}
\end{align}
Therefore, the Euclidean gradient of the weighted sum rate function $\nabla f_{1}(\boldsymbol{\theta})$ is given by
\begin{equation}
\begin{split}
&\nabla f_{1}(\boldsymbol{\theta})=\\
&\sum_{k=1}^{K}w_{k}\frac{1}{\ln 2}\frac{\boldsymbol{\theta}\mathrm{diag}(\mathbf{h}_{r,l,k}^{*})\mathbf{G}_{l}(\sum_{i=1}^{K}p_{i}\mathbf{w}_{i}\mathbf{w}_{i}^{H})\mathbf{G}_{l}^{H}\mathrm{diag}(\mathbf{h}_{r,l,k})^{T}}{\sigma^{2}+\sum_{i=1}^{K}|\boldsymbol{\theta}\mathrm{diag}(\mathbf{h}_{r,l,k}^{*})\mathbf{G}_{l}\mathbf{w}_{i}|^{2}}\nonumber\\
&-w_{k}\frac{1}{\ln 2}\frac{\boldsymbol{\theta}\mathrm{diag}(\mathbf{h}_{r,l,k}^{*})\mathbf{G}_{l}(\sum_{i\neq k}^{K}p_{i}\mathbf{w}_{i}\mathbf{w}_{i}^{H})\mathbf{G}_{l}^{H}\mathrm{diag}(\mathbf{h}_{r,l,k})^{T}}{\sigma^{2}+\sum_{i\neq k}^{K}|\boldsymbol{\theta}\mathrm{diag}(\mathbf{h}_{r,l,k}^{*})\mathbf{G}_{l}\mathbf{w}_{i}|^{2}},\label{16}
\end{split}
\end{equation}
and $\nabla f_{2}(\mathbf{W})$ is given in~(\ref{17}) on the next page.
\newcounter{mytempeqncnt}
\begin{figure*}[!t]
\normalsize
\setcounter{mytempeqncnt}{\value{equation}}
\setcounter{equation}{16}
\begin{equation}
\begin{split}
&\nabla f_{2}(\mathbf{W})=\\
&\left[
 \begin{matrix}
\frac{w_{1}p_{1}\mathbf{w}_{1}^{H}\mathbf{G}_{l}^{H}\mathrm{diag}(\mathbf{h}_{r,l,1})^{T}\boldsymbol{\theta}^{H}\boldsymbol{\theta}\mathrm{diag}(\mathbf{h}_{r,l,1}^{*})\mathbf{G}_{l}}{\ln 2(\sigma^{2}+\sum_{i=1}^{K}|\boldsymbol{\theta}\mathrm{diag}(\mathbf{h}_{r,l,1}^{*})\mathbf{G}_{l}\mathbf{w}_{i}|^{2})}-\sum_{j\neq 1}^{K}
\frac{w_{j}p_{1}p_{j}\mathbf{w}_{1}^{H}\mathbf{G}_{l}^{H}\mathrm{diag}(\mathbf{h}_{r,l,j})^{T}\boldsymbol{\theta}^{H}\boldsymbol{\theta}\mathrm{diag}(\mathbf{h}_{r,l,j}^{*})\mathbf{G}_{l}|(\boldsymbol{\theta}\mathrm{diag}(\mathbf{h}_{r,l,j}^{*})\mathbf{G}_{l})\mathbf{w}_{j}|^{2}}{\ln 2(\sigma^{2}+\sum_{i=1}^{K}|\boldsymbol{\theta}\mathrm{diag}(\mathbf{h}_{r,l,j}^{*})\mathbf{G}_{l}\mathbf{w}_{i}|^{2}p_{i})(\sigma^{2}+\sum_{i\neq j}^{K}|\boldsymbol{\theta}\mathrm{diag}(\mathbf{h}_{r,l,j}^{*})\mathbf{G}_{l}\mathbf{w}_{j}|^{2}p_{i})}
\\
\vdots\\
\frac{w_{K}p_{K}\mathbf{w}_{K}^{H}\mathbf{G}_{l}^{H}\mathrm{diag}(\mathbf{h}_{r,l,K})^{T}\boldsymbol{\theta}^{H}\boldsymbol{\theta}\mathrm{diag}(\mathbf{h}_{r,l,K}^{*})\mathbf{G}_{l}}{\ln 2(\sigma^{2}+\sum_{i=1}^{K}|\boldsymbol{\theta}\mathrm{diag}(\mathbf{h}_{r,l,K}^{*})\mathbf{G}_{l}\mathbf{w}_{i}|^{2})}-\sum_{j\neq K}^{K}
\frac{w_{j}p_{K}p_{j}\mathbf{w}_{K}^{H}\mathbf{G}_{l}^{H}\mathrm{diag}(\mathbf{h}_{r,l,j})^{T}\boldsymbol{\theta}^{H}\mathrm{diag}(\mathbf{h}_{r,l,j}^{*})\mathbf{G}_{l}|(\boldsymbol{\theta}\mathrm{diag}(\mathbf{h}_{r,l,j}^{*})\mathbf{G}_{l})\mathbf{w}_{j}|^{2}}{\ln 2(\sigma^{2}+\sum_{i=1}^{K}|\boldsymbol{\theta}\mathrm{diag}(\mathbf{h}_{r,l,j}^{*})\mathbf{G}_{l}\mathbf{w}_{i}|^{2}p_{i})(\sigma^{2}+\sum_{i\neq j}^{K}|\boldsymbol{\theta}\mathrm{diag}(\mathbf{h}_{r,l,j}^{*})\mathbf{G}_{l}\mathbf{w}_{j}|^{2}p_{i})}\\
  \end{matrix}
  \right].
\end{split}~\label{17}
\end{equation}
\setcounter{equation}{\value{mytempeqncnt}}
\hrulefill
\vspace*{4pt}
\end{figure*}
\setcounter{equation}{17}

We reformulate the constraint~(\ref{12b}) and~(\ref{12c}) as two manifolds and define the manifold  $\mathcal{M}$ as
\begin{equation}
\begin{split}
\mathcal{M}=\{\boldsymbol{\theta}\in\mathbb{C}^{MK\times 1}:|\theta_{i}|=1,i=1,\cdots,MK\},
\end{split}
\end{equation}
where $\mathcal{M}$ is called the Riemannian manifold~\cite{absil2009optimization}. The manifold $\mathcal{M}$ corresponds to the unit modulus constraint. Then, we note that the amplitude constraint of active beamforming defines an Oblique manifold $\mathcal{N}$ which can be characterized as
\begin{equation}
\begin{split}
\mathcal{N}=\{\mathbf{W}\in\mathbb{C}^{K\times N}|\mathbf{I}_{K\times K}\circ(\mathbf{W}\mathbf{W}^{H})=\mathbf{I}_{K\times K}\}.
\end{split}
\end{equation}

The tangent space to  $\mathcal{M}$ at point $\boldsymbol{\theta}$ is denoted as $\mathcal{T}_{\boldsymbol{\theta}}{\mathcal{M}}$. Given a weighted sum rate cost function $f_{1}(\boldsymbol{\theta})$ on a Riemannian manifold $\mathcal{M}$, the directional derivative of $f_{1}(\boldsymbol{\theta})$ along $\boldsymbol{\omega}\in\mathcal{T}_{\boldsymbol{\theta}}{\mathcal{M}}$ can be denoted by $D_{\boldsymbol{\omega}}f_{1}(\boldsymbol{\theta})$. $\mathrm{grad}_{\mathcal{M}}f_{1}(\boldsymbol{\theta})$ denotes the gradient of $f_{1}(\boldsymbol{\theta})$ at $\boldsymbol{\theta}$. According to~\cite{absil2009optimization}, the elements of $\mathcal{T}_{\boldsymbol{\theta}}{\mathcal{M}}$ satisfy
\begin{equation}
\begin{split}
D_{\boldsymbol{\omega}}f_{1}(\boldsymbol{\theta})=\boldsymbol{\omega}\circ\mathrm{grad}_{\mathcal{M}}f_{1}(\boldsymbol{\theta}), \forall \boldsymbol{\omega} \in\mathcal{T}_{\boldsymbol{\theta}}{\mathcal{M}}.
\end{split}
\end{equation}
Similarly, the elements of $\mathcal{T}_{\mathbf{W}}{\mathcal{N}}$ also have
\begin{equation}
\begin{split}
D_{\boldsymbol{\psi}}f_{2}(\mathbf{W})=\boldsymbol{\psi}\circ\mathrm{grad}_{\mathcal{N}}f_{2}(\mathbf{W}),\forall \boldsymbol{\psi}\in\mathcal{T}_{\mathbf{W}}{\mathcal{N}},
\end{split}
\end{equation}
where $\mathcal{T}_{\mathbf{W}}{\mathcal{N}}$ is the tangent space to $\mathcal{N}$ at the point $\mathbf{W}$. The gradient of $f_{1}(\boldsymbol{\theta})$ and $f_{2}(\mathbf{W})$ on the complex Riemannian manifold and the complex Oblique manifold are derived here.  The normal space to $\mathcal{M}$ and $\mathcal{N}$ at the point $\boldsymbol{\theta}$ and $\mathbf{W}$ are denoted as $\widetilde{\mathcal{N}}_{\boldsymbol{\theta}}{\mathcal{M}}$ and $\widetilde{\mathcal{N}}_{\mathbf{W}}{\mathcal{N}}$, respectively. The gradient of weighted sum rate cost function $f_{1}(\boldsymbol{\theta})$ and $f_{2}(\mathbf{W})$ on the manifold $\mathcal{M}$ and $\mathcal{N}$ can be respectively expressed as
\begin{align}
\mathrm{grad}_{\mathcal{M}}f_{1}(\boldsymbol{\theta})&=\nabla f_{1}(\boldsymbol{\theta})-\mathfrak{R}[\nabla f_{1}(\boldsymbol{\theta})\circ\boldsymbol{\theta}^{*}]\circ\boldsymbol{\theta},\\
\mathrm{grad}_{\mathcal{N}}f_{2}(\mathbf{W})&=\nabla f_{2}(\mathbf{W})-
(\mathbf{I}_{M}\circ\mathfrak{R}\{\mathbf{W}(\nabla f_{2}(\mathbf{W}))^{H}\})\mathbf{W}.
\end{align}

According to proofs in~\cite{absil2009optimization, absil2006joint}, projection function $P_{\boldsymbol{\theta}}$ and $P_{\mathbf{W}}$ can be denoted as
\begin{align}
P_{\boldsymbol{\theta}}(\nabla f_{1}(\boldsymbol{\theta}))&=\nabla f_{1}(\boldsymbol{\theta})-\mathfrak{R}[\nabla f_{1}(\boldsymbol{\theta})\circ\boldsymbol{\theta}^{*}]\circ\boldsymbol{\theta},\\
P_{\mathbf{W}}(\nabla f_{2}(\mathbf{W}))&=\nabla f_{2}(\mathbf{W})-
(\mathbf{I}_{M}\circ\mathfrak{R}\{\mathbf{W}(\nabla f_{2}(\mathbf{W}))^{H}\})\mathbf{W}.
\end{align}

In order to stay on the manifold, one can also apply the concept of the retraction~\cite{absil2009optimization}. Given the search step size $\alpha$ and $\beta$, the search direction $\mathbf{d}_{\alpha}$ and $\mathbf{d}_{\beta}$,  retractions on $\mathcal{M}$ and $\mathcal{N}$ are expressed as
\begin{equation}
\begin{split}
\mathrm{Ret}_{\boldsymbol{\theta}}(\alpha\mathbf{d}_{\alpha})=
\left [
\begin{matrix}
\frac{\theta_{1}+\alpha d_{\alpha,1}}{|\theta_{1}+\alpha d_{\alpha,1}|},\cdots ,\frac{\theta_{LM}+\alpha d_{\alpha,LM}}{|\theta_{LM}+\alpha d_{\alpha,LM}|}
\end{matrix}
\right ]^{T},
\end{split}\label{26}
\end{equation}
\begin{equation}
\begin{split}
&\mathrm{Ret}_{\mathbf{W}}(\beta\mathbf{d}_{\beta})=\\
&\left [
\begin{matrix}
\frac{W_{11}+\beta d_{\beta,11}}{\sqrt{\sum_{i=1}^{K}|W_{1i}+\beta d_{\beta,1i}|^{2}}}&\cdots &\frac{W_{1K}+\beta d_{\beta,1K}}{\sqrt{\sum_{i=1}^{K}|W_{1i}+\beta d_{\beta,1i}|^{2}}}\\
\vdots&\ddots&\vdots\\
\frac{W_{N1}+\beta d_{\beta,N1}}{\sqrt{\sum_{i=1}^{K}|W_{Ni}+\beta d_{\beta,Ni}|^{2}}}&\cdots &\frac{W_{NK}+\beta d_{\beta,NK}}{\sqrt{\sum_{i=1}^{K}|W_{Ni}+\beta d_{\beta,Ni}|^{2}}}
\end{matrix}
\right ].~\label{27}
\end{split}
\end{equation}

Next, we briefly recall the conjugate gradient algorithm and extend this algorithm to the manifold. Then, the algorithms for active/passive beamforming design under the two manifolds are derived. The update conjugate direction is used to search a maximum of the function $f_{1}(\boldsymbol{\theta})$ and $f_{2}(\mathbf{W})$. They are given by
\begin{align}
\mathbf{d}^{(t+1)}_{\alpha}&=\mathbf{g}_{\alpha}^{(t+1)}+\lambda_{\alpha}^{(t+1)}P_{\boldsymbol{\theta}^{(t+1)}}(\mathbf{d}^{t}_{\alpha}),\label{28}\\
\mathbf{d}^{(t+1)}_{\beta}&=\mathbf{g}_{\beta}^{(t+1)}+\lambda_{\beta}^{(t+1)}P_{\mathbf{W}^{(t+1)}}(\mathbf{d}^{t}_{\beta}),~\label{29}
\end{align}
where $\lambda_{\alpha}^{(t+1)}$ and $\lambda_{\beta}^{(t+1)}$ are Polak-Ribiere parameter in the $t+1$-th iteration~\cite{absil2009optimization}. They can be computed by
\begin{align}
\lambda_{\alpha}^{(t+1)}&=\frac{(\mathbf{g}_{\alpha}^{(t+1)})^{H}(\mathbf{g}_{\alpha}^{(t+1)}-P_{\boldsymbol{\theta}^{(t+1)}}(\mathbf{g}_{\alpha}^{(t)}))}{\|P_{\boldsymbol{\theta}^{(t+1)}}(\mathbf{g}_{\alpha}^{(t)})\|^{2}},\label{30}\\
\lambda_{\beta}^{(t+1)}&=\frac{(\mathbf{\beta}_{\alpha}^{(t+1)})^{H}(\mathbf{g}_{\beta}^{(t+1)}-P_{\boldsymbol{\theta}^{(t+1)}}(\mathbf{g}_{\beta}^{(t)}))}{\|P_{\boldsymbol{\theta}^{(t+1)}}(\mathbf{g}_{\beta}^{(t)})\|^{2}}.\label{31}
\end{align}

$\mathbf{g}_{\alpha}^{(t+1)}$ and $\mathbf{g}_{\beta}^{(t+1)}$ are gradient update in $t+1$-th iteration, which can be computed as
\begin{align}
\mathbf{g}^{(t+1)}_{\alpha}&=P_{\boldsymbol{\theta}^{(t+1)}}(\nabla f_{1}(\boldsymbol{\theta}^{(t+1)})),\label{32}\\
\mathbf{g}^{(t+1)}_{\beta}&=P_{\mathbf{W}^{(t+1)}}(\nabla f_{2}(\mathbf{W}^{(t+1)})),\label{33}
\end{align}

\begin{algorithm}[t]
\caption{Alternative Manifold Optimization } 
\label{algo-1}
\hspace*{0.02in}{\bf Input:} $\mathbf{p}^{(0)}$, $\boldsymbol{\theta}^{(0)}$, $\mathbf{W}^{(0)}$, $t=0$ and $f(\boldsymbol{\theta},\mathbf{p},\mathbf{W})^{(-1)}=0$\\
\hspace*{0.02in}{\bf Repeat:}\\
With the current $\mathbf{p}^{(t)}$ and $\mathbf{W}^{(t)}$, update $\boldsymbol{\theta}^{(t+1)}$\\
\hspace*{0.02in}{\bf Repeat:}\\
Computing the Armijo search step size $\alpha^{(t)}$\cite{absil2009optimization}\\
Update the Riemannian gradient based on (\ref{32})\\
Calculate Polak Ribiere parameter $\lambda_{\alpha}^{(t+1)}$ based on (\ref{30})\\
Determine search direction $\mathbf{d}^{(t+1)}_{\alpha}$ based on (\ref{28})\\
Computing $\boldsymbol{\theta}^{(t+1)}$ according to (\ref{34})\\
\hspace*{0.02in}{\bf Until:}
$|f(\boldsymbol{\theta}^{(t+1)})-f(\boldsymbol{\theta}^{(t)})|<\upsilon$\\
With the current $\mathbf{p}^{(t)}$ and $\boldsymbol{\theta}^{(t+1)}$, update $\mathbf{W}^{(t+1)}$\\
\hspace*{0.02in}{\bf Repeat:}\\
Computing the Armijo search step size $\beta^{(t)}$\cite{absil2009optimization}\\
Update the Riemannian gradient based (\ref{33})\\
Calculate Polak Ribiere parameter $\lambda_{\beta}^{(t+1)}$ based on (\ref{31})\\
Determine search direction $\mathbf{W}^{(t+1)}$ based on (\ref{29})\\
Find $\mathbf{W}^{(t+1)}$ according to (\ref{35})\\
\hspace*{0.02in}{\bf Until:}
$|f(\mathbf{W}^{(t+1)})-f(\mathbf{W}^{(t)})|<\nu$\\
\hspace*{0.02in}{\bf Repeat:}\\
Update $\mathbf{p}^{(t+1)}$ by using GP algorithm\cite{chiang2008power}\\
\hspace*{0.02in}{\bf Until:}$|f(\boldsymbol{\theta}^{(t+1)},\mathbf{W}^{(t+1)},\mathbf{p}^{(t+1)})-f(\boldsymbol{\theta}^{(t)},$$\mathbf{W}^{(t)},\mathbf{p}^{(t)})|<\zeta$\\
\hspace*{0.02in}{\bf Output:} 
$\boldsymbol{\theta}^{(t+1)}$, $\mathbf{W}^{(t+1)}$, $\mathbf{p}^{(t+1)}$ 
\end{algorithm}

The step size $\alpha^{(t+1)}$ and $\beta^{(t+1)}$ can be chosen by a line-search algorithm, however, the calculation is very expensive. Therefore, in this paper, we adopt the Armijo backtracking line search to determine the step size $\alpha^{(t+1)}$ and $\beta^{(t+1)}$, the method is given in  Definition 4.2.2 of~\cite{absil2009optimization}.

The update points are given by using retraction operation in (\ref{26}), (\ref{27}) and they are expressed as
\begin{equation}
\begin{split}
\boldsymbol{\theta}^{(t+1)}=\mathrm{Ret}_{\boldsymbol{\theta}^{(t)}}(\alpha^{(t+1)}\mathbf{d}_{\alpha}^{(t+1)}),
\end{split}\label{34}
\end{equation}
\begin{equation}
\begin{split}
\mathbf{W}^{(t+1)}=\mathrm{Ret}_{\mathbf{W}^{(t)}}(\beta^{(t+1)}\mathbf{d}_{\beta}^{(t+1)}).
\end{split}\label{35}
\end{equation}
The alternating maximization algorithm based manifold is summarized in Algorithm~\ref{algo-1}.

\subsection{Convergence Analysis}
In this section, we discuss the convergence for  Algorithm~\ref{algo-1}
and analyze its computational complexity. Note that the power allocation algorithm has a solution that satisfies the Karush-Kuhn-Tucker (KKT) conditions~\cite{boyd2004convex}. Thus, given a passive/active beamforming matrix, the optimization step of the allocation power matrix always ensures the increase of the objective function~\cite{boyd2004convex}. Hence, the convergence of each algorithm depends on its optimization step for active/passive beamforming design. According to Theorem 4.3.1 in~\cite{absil2009optimization}, the algorithm using the manifold optimization is guaranteed to convergence to the point where the gradient of the objective function is $0$~\cite{absil2009optimization}. Therefore, each step of the whole alternating Algorithm~\ref{algo-1} ensures the increase of the objective function and obtains a local optimal solution in each iteration. \vspace{-5pt}

\subsection{Computational complexity}
Complexity in each inner iteration includes the seven parts:

\begin{itemize}
\item
Computation of the power allocation:
according to \cite{chiang2008power}, the complexity of the power allocation algorithm is $\mathcal{O}(N_{0}\max\{K^{3},F\})$, where $N_{0}$ is the number of iteration for convergence of the algorithm and $F$ is the first and second derivatives of the objective functions.
\item
Computation of the conjugate gradient of active/passive beamforming: according to ~(\ref{16}) and~(\ref{17}), the total complexity in computing the gradient is $6KNLM+(2K^{2}-1)N^{2}$ and $(5K^{2}+2K-4)LMN$, respectively.

\item
The projection and retraction of active/passive beamforming operations:
the complexity of the  retraction  operations is $LM$ and $KN$. The complexity of projection operations is $2LM$ and $2KN$, respectively.

\item
Line search for active/passive beamforming: 
the complexity of the Armijo backtracking line search is $6LM$ and $KN^{2}+6KN$, respectively.
 
\end{itemize}
Therefore, the total complexity of Algorithm~\ref{algo-1} is $\mathcal{O}(N_{0}\max\{K^{3},F\})+N_{1}(6KNLM+(2K^{2}-1)N^{2}+9LM)+N_{2}((5K^{2}+2K-4)LMN+9KN+KN^{2})$.

\section{Numerical Results}\label{VI}

\begin{figure}[htbp]
\centering
\includegraphics[height=1.0in]{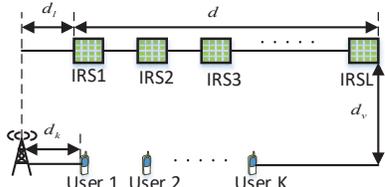}
\vspace{-7pt}\caption{Simulation parameters of Multiple-IRS assist mmWave communication system setup.\vspace{-20pt}}
\label{fig-2}
\end{figure}

In this section, we present the numerical results. The locations of the BS, IRSs, and users are shown in Fig.~\ref{fig-2}. According to~\cite{akdeniz2014millimeter}, the path loss is taken as
\begin{equation}
PL(d)[dB]=\alpha+\beta\log_{10}(d)+\xi,~~~~ \xi\sim\mathcal{CN}(0,\sigma^{2}_{\xi}),
\end{equation}
where $\alpha=61.4dB$ and $\beta=20$. 
The number of antennas of BS $N=32$ and BS is located in the origin point. 
The communication system's layout is illustrated in Fig.~\ref{fig-2}, where $L$ IRSs are equally spaced on a straight line which is in parallel with the line connecting the BS and the user. The horizontal distance between the BS and the first IRS is set to $d_{l}=11m$ and the vertical distance is set to $d_{v}=1m$.
The K users are distributed in a line between $0m$ and $40m$ from the BS. For ease of simulation, we set $K = 2, 4, 6$ with an interval of $5$m among each user, while the distance between the BS and User1 is set as $d_{k}=5m$. Let $L$ IRSs locate in a line between $11m$ and $50m$ with a uniform distribution. 
In addition, the transmit power is set as $P=30dBm$ and the noise power $P_{n}=-85dBm$ \cite{akdeniz2014millimeter}.\vspace{-10pt}

\begin{figure}[h]
\hspace{-30pt}
\includegraphics[height=2.5in]{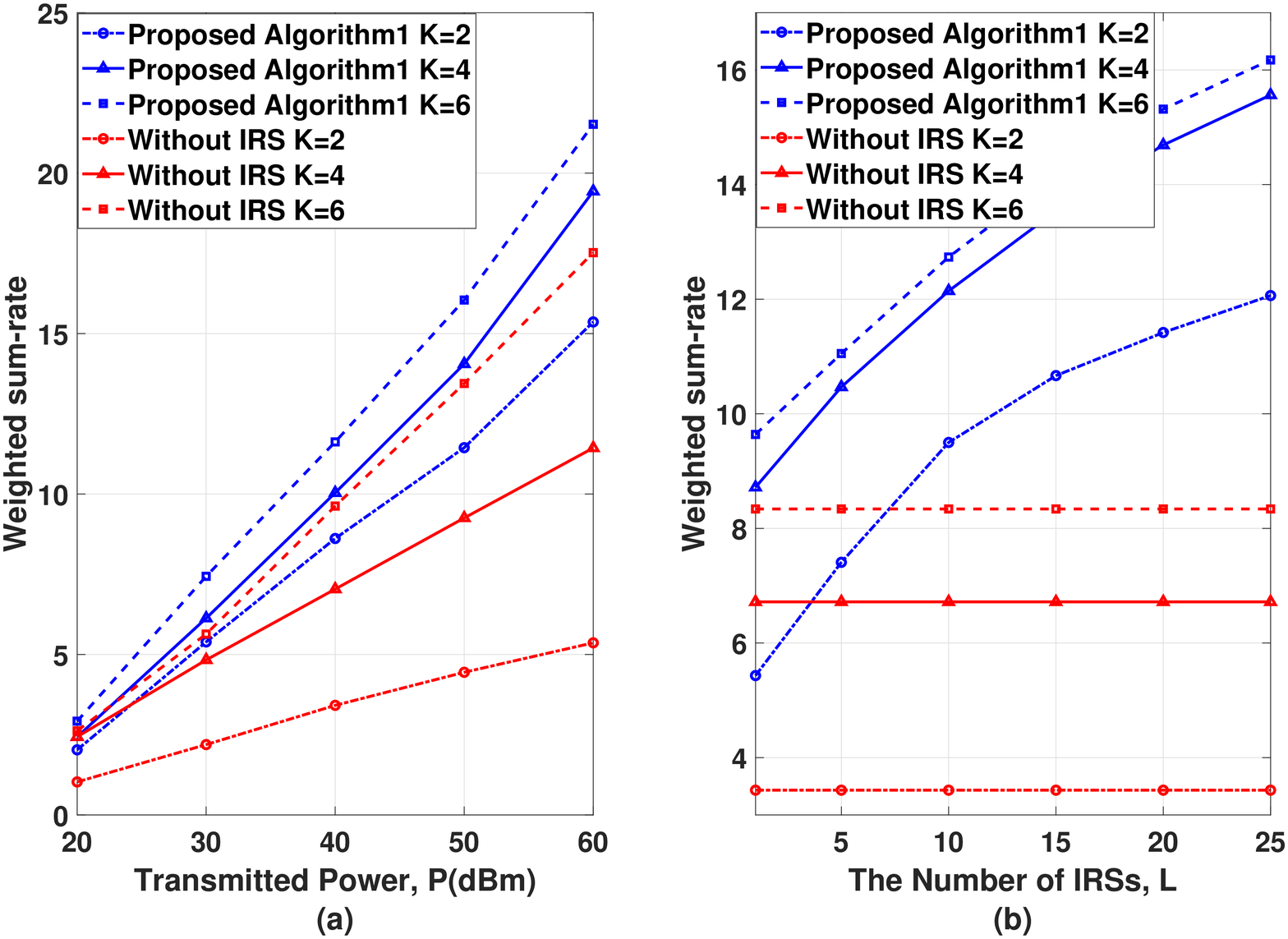}\vspace{-7pt}
         \caption{The sum rate vs transmit power $P$ in (a), and the sum rate vs the number of IRSs in (b).\vspace{-10pt}}
         \label{fig:log}
\end{figure}

Fig.~\ref{fig:log}(a) plots the sum-rate of our proposed algorithms vs. the transmitted power $P$ between the BS and the user, where the number of IRSs is $L=2$. It can be observed that the IRS-assisted system can help substantially improve the sum rate. Moreover, in Fig.~\ref{fig:log}(b), we plot the sum rate versus the number of IRS units where each IRS equipped with $M=20$ reflecting elements and the BS with $N=32$ antennas. The number of IRSs $L$ gradually increases from $1$ to $25$. Clearly, all of multi-user scenarios exhibit the same upward trend.

\begin{figure}[!t]
\centering
\includegraphics[scale=0.4]{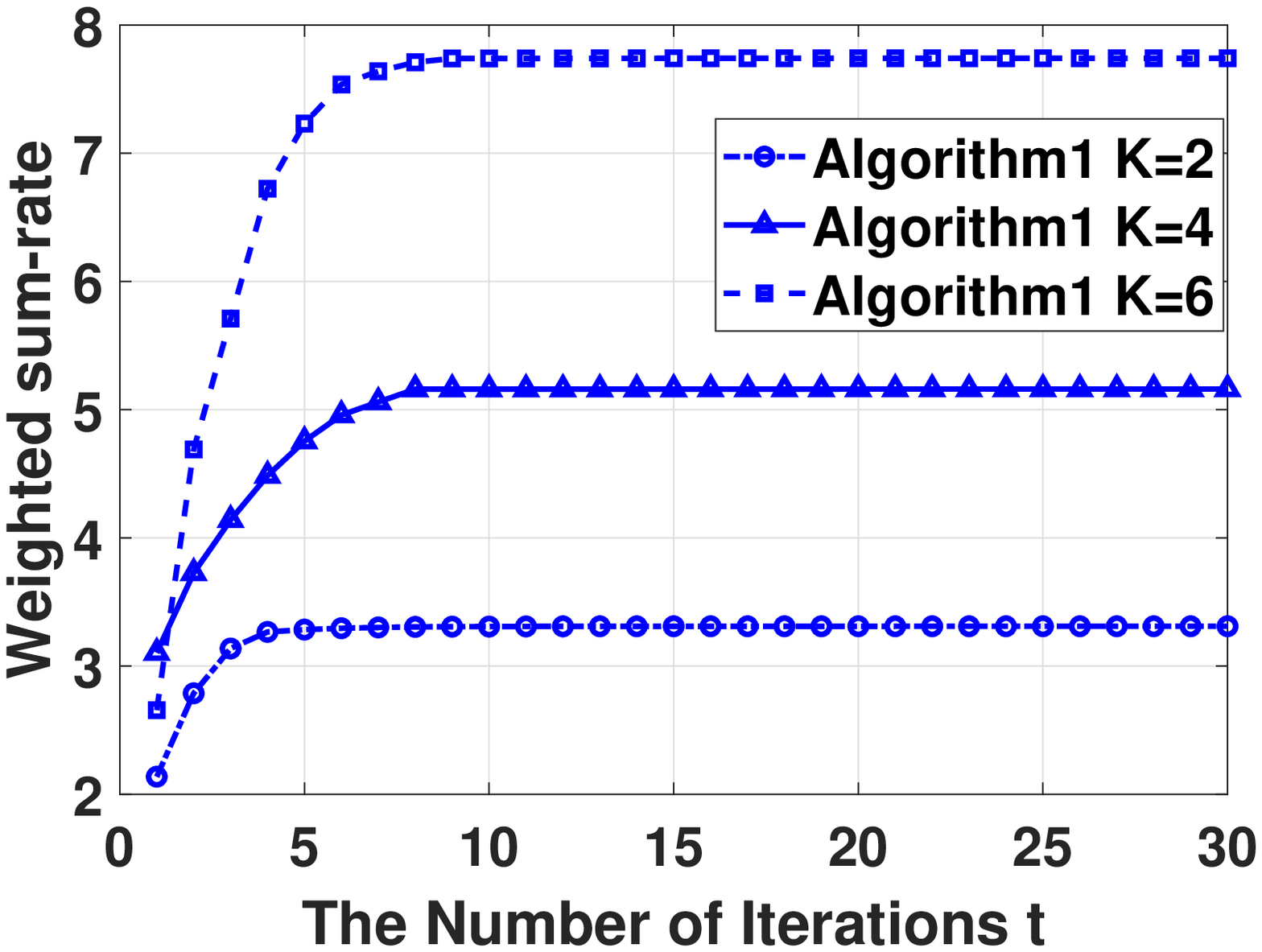}\vspace{-7pt}
\caption{Convergence behavior of the proposed Algorithm~\ref{algo-1} with $M=20$ and $N=20$.\vspace{-15pt}}
\label{fig-4}
\end{figure}
In Fig.~\ref{fig-4}, we show the convergence trend of the Algorithm~\ref{algo-1}. From the simulation results, the convergence of our proposed algorithm is confirmed in multiple cases, i.e., $K=2,4,6$. Our proposed algorithms converge after about 6 iterations. Meanwhile, we observe that as the number of users increases, the sum rate also increases after convergence. 

\begin{figure}[!t]
\centering
\includegraphics[scale=0.3]{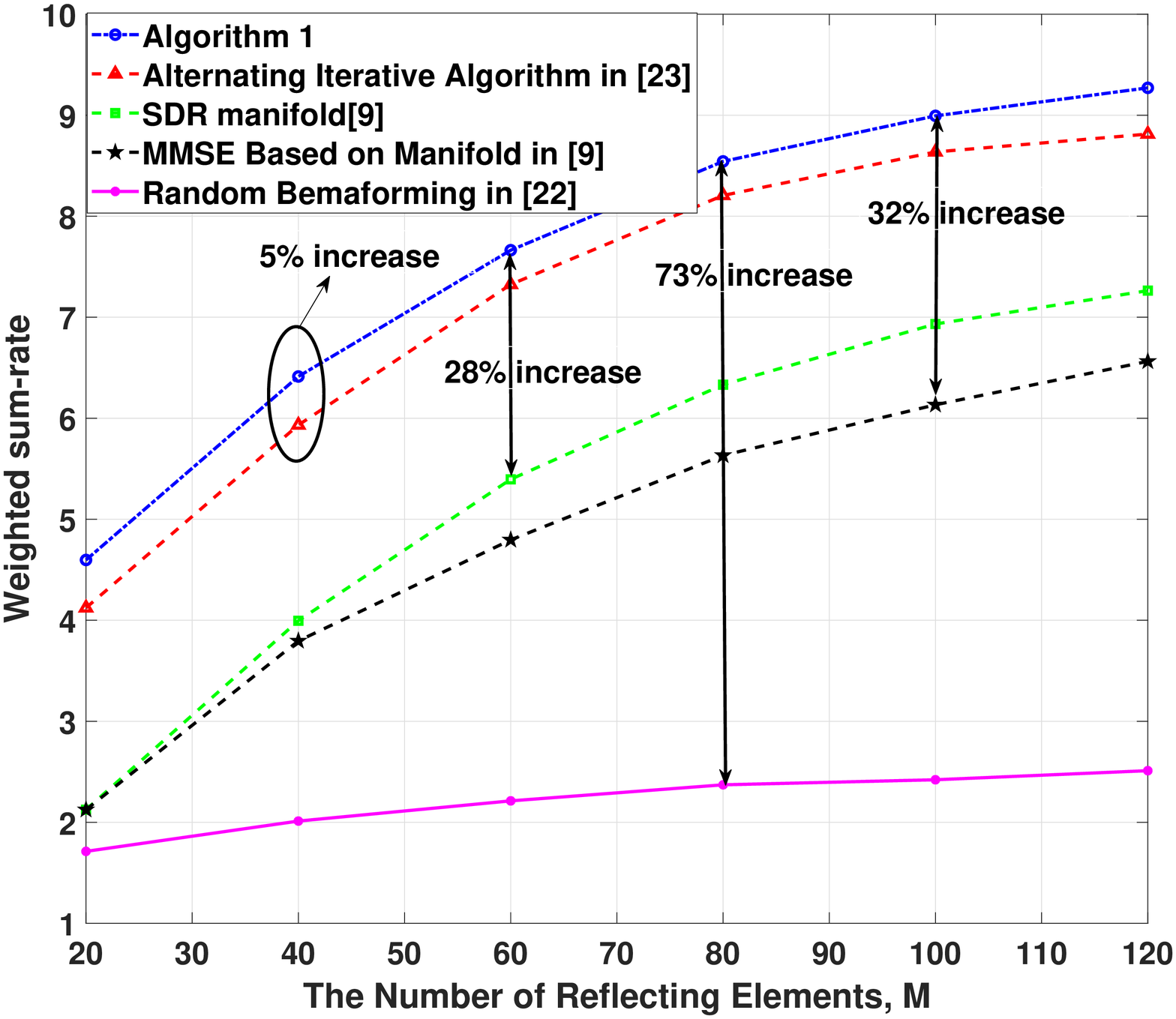}\vspace{-7pt}
\caption{Convergence behavior of the proposed Algorithm~\ref{algo-1} with $M=20$ and $N=20$.\vspace{-20pt}}
\label{fig-5}
\end{figure}
In Fig.~\ref{fig-5}, we investigate the sum-rate achieved by different algorithms when $N=32$, $L=2$ and $K=6$. As shown in Fig.~\ref{fig-5}, with the number of reflecting elements increasing from 
$20$ to $120$, the existing random beamforming algorithm~\cite{lee2015randomly} achieves a significantly lower sum-rate than the proposed algorithm. Our proposed algorithm achieves the best performance over the IRS size range in consideration. From Fig.~5, we find the proposed Algorithm~\ref{algo-1} achieves substantial performance gain. \vspace{-4pt}

\section{Conclusion}\label{VII}
In this paper, we investigated the joint power allocation and beamforming design to maximize the weighted sum-rate in an IRS-assisted mmWave communication system. The power allocation problem is resolved using the geometric programming (GP) algorithm. Then, we fix the power allocation matrix, and the considered beamforming design problem is viewed as a constrained optimization problem over two manifolds.  By exploiting the principle of alternating optimization, we derive the gradient of the objective function on the two manifolds. Then, we propose to use a conjugate gradient method to search the active/passive beamforming matrix. Numerical results show that compared with existing schemes, the proposed algorithm improves the weighted sum-rate of the system and decreases the computational cost. \vspace{-7pt}



%

\appendices
\ifCLASSOPTIONcaptionsoff
  \newpage
\fi







\bibliographystyle{abbrv}
\bibliography{related}
\end{document}